\documentstyle[amssymb,floats,aps]{revtex}

\begin{document}

\twocolumn[\hsize\textwidth\columnwidth\hsize\csname  
@twocolumnfalse\endcsname

\title{Analogy Between Periodically Driven Josephson ac Effect and Quantum Hall Effect}

\author{T.C. Wang}

\address{
Department of Electrophysics, National Chiao-Tung University\\
Hsinchu, Taiwan}

\maketitle
\begin{abstract}
The hydrodynamical phase-slippage geometry and a semi-classical
Josephson-RCSJ formulism are combined to make an analogy between the periodically
driven Josephson ac effect and the quantum Hall effect.
Both of these two macroscopic quantum effects are postulated to be originated from  the
one-dimensional linear vortex-electron relative motion. The mechanisms of the quantized Hall
resistance of IQHE and FQHE are unified under the phase-locking dynamics of a
Josephson-like oscillation which is periodically driven by the electron wave of the coherent
Hall current, with the time-averaged frequency ratio playing the role of the filling factor
reciprocal.\\
  Supporting experimental evidences for this scenario as well as necessary future investigations are
remarked.\\
\end{abstract}

\vskip2pc]

\narrowtext

When the applied current is larger than a critical value, the Josephson
junction develops an ac current along with a voltage drop across the
barrier. This is known as the ac Josephson effect\cite{JJ}. If the junction
is further exposed to a microwave irradiation with an angular frequency $%
\omega _{r.f}$, the current-voltage characteristic curve shows Shapiro
voltage steps\cite{ISH}\cite{FSH}: 
\begin{equation}
V=\alpha \frac{\hbar }{2e}\omega _{r.f},  \label{alpha}
\end{equation}

\noindent where $2e$ denotes the electric charge of the Cooper pair, $V$
represents the voltage drop across the junction, the step index $\alpha $
can be an integer or a fractional number and the Planck constant $\hslash
=h/2\pi $. According to such a voltage-frequency relation, the r.f-biased ac
Josephson effect provides a metrological quantum standard of the voltage.

To realize the Shapiro steps, one starts from the Josephson relations\cite
{JJ}\cite{feynman} 
\begin{equation}
V\ =\frac{\hbar }{2e}\frac{d\varphi }{dt}  \label{VJ}
\end{equation}
where $\varphi $ represents the electronic phase difference across the
junction, and 
\begin{equation}
I_{J}\ =I_{c}\sin \varphi  \label{IJ}
\end{equation}
where $I_{J}$ is the Josephson supercurrent, $I_{c}$ is the critical
current, and then solves the semi-classical equation describing the current
conservation across the junction. A discrete time computing of the
well-known, r-f biased, Resistively and Capacitively Shunted Josephson
Junction (RCSJ\cite{stewart}\cite{mccumber}) model shows\cite{kautz} the
main features of the I-V curve which includes various typical properties of
a (2+1)-dimensional nonlinear dynamical system such as deterministic chaos,
quasiperiodic and the phase-locking behavior. The Shapiro step index $\alpha 
$ can be shown as an integral, or a fractional, winding number\cite{bohr},
namely, time averaged frequency ratio $<\Delta \varphi /\Delta t>/\omega
_{r.f}$, in the phase-locked stationary states of the dynamical system.

Serving the metrology as another macroscopic quantum standard of a classical
quantity, the transverse Hall resistance $R_{H}$ in the Quantum Hall effect
(QHE)\cite{prange}\cite{chakraborty} can be quantized in units of $h/e^{2}$: 
\begin{equation}
R_{H}\ =\frac{1}{\nu }\frac{h}{e^{2}},  \label{nu}
\end{equation}
once the longitudinal Hall current simultaneously becomes perfectly
conducting. In (\ref{nu}) $\nu $ is an integer for the integer quantum Hall
effect (IQHE)\cite{iqhe} or a fractional number for the fractional quantum
Hall effect (FQHE)\cite{fqhe}.

Interestingly, we see from a dimensional check that the quantized Hall
resistance can be expressed as $h/e^{2}$ multiplied by the frequency ratio
of a Josephson-like oscillation to an electron wave: 
\begin{equation}
R_{H}=\frac{V_{H}}{I_{H}}[=\frac{\left( \hbar /e\right) \omega _{H}}{ef_{e}}%
=(\frac{\omega _{H}}{\omega _{e}})(\frac{h}{e^{2}})],  \label{dimension}
\end{equation}
where $V_{H}$ and $I_{H}$ is the Hall voltage and the Hall current
respectively, $[...]$ denotes dimensional equalities, therein $\omega _{H}$
implies the angular frequency of a presumable single charge Josephson
oscillation and $f_{e}(=\omega _{e}/2\pi )$ represents the frequency of a
coherent electron wave. In the following work, we would like to answer the
inspiring question: Is it possible to connect these two effects in a unified
framework which solves the IQHE and FQHE by a common dynamical theory?

Compared to the simplicity the semi-classical phase-locking picture offered
in describing both the integer and fractional Shapiro steps in a unified
scheme, up to date quantum mechanical approaches from IQHE to FQHE appear to
be more complicated. Without loosing the basic common quantum-mechanical
features of those treatments, we take the Gauge Invariance picture for
example to review the geometry, based on which the boundary condition for
QHE was given. If a closed integral path $\oint dl$ is taken to be a loop of
electron current flow, as that suggested by Laughlin's ribbon loop\cite
{laughlin}, the quantization behavior of the charge carriers in IQHE can be
attributed to the single-electron gauge invariance condition

\begin{equation}
\oint {\bf A\cdot }d{\bf l}=n\Phi _{0},  \label{gi}
\end{equation}
where ${\bf A}$ is the vector potential of the gauge field threading the
looping current, ${\bf l}$ implies the position of the circulating electrons
and $\Phi _{0}$ denotes the gauge flux quantum $h/e$. This current loop
geometry provided a periodic boundary condition by which the phase increment
of the electrons during each cycle as well as crossing from one edge of the
ribbon to the other was quantum-mechanically restricted to be $2\pi $, and
the factor $\nu $ of (\ref{nu}) for IQHE was shown to be the number of
charges involved in the cycling motion. According to this imaginary
geometry, the FQHE can be realized by the introduction of a fractional
charge state\cite{anderson} for an interactive electron fluid within the
many-body theoretical framework. On the other hand, as to be illustrated
below, the vortex motion scenario figured by P.W. Anderson\cite{p-slip} for
the ac Josephson effect provides another possible, and even more realistic,
geometric relation between the Hall current and voltage drop to show the
quantized Hall resistance.

In the phase-slippage picture, a quantization condition subjected to the
Josephson voltage relation (\ref{VJ}) was adopted in almost a same
mathematical form compared with (\ref{gi}): 
\begin{equation}
\oint {\bf \nabla }\phi \cdot d{\bf s}=2n\pi ,  \label{gij}
\end{equation}
only that the integral path $\oint ds$ was notably taken around an isolated
Josephson vortex core as shown in Fig.1(a) instead of a trajectory of
circulating current, herein $\phi $ denoted the electronic phase. It should
be noted that the crucial underlying physical difference between these two
pictures is that the former provides a quantization criterion for the single
electron particle surrounding a flux while the latter has shown by itself a
quantization of the vortex existing in the electronic fluid. To recall a
real image of the relative motion between electron fluid and vortices, one
considers a thin layer Josephson junction lying on the $x-z$ plane with the
voltage drop developing along the $y$ direction and the vortex moving along
the $x$ direction, see Fig.1(b). If the electronic phase difference between
side $1$ and side $2$ of the junction barrier, $\varphi _{12}(=\phi
_{1}-\phi _{2})$, at a certain $x$ position is measured, one finds that it
gains $2\pi $ increment as each traveling vortex passes through that $x$
position. An analogy of quantum oscillation can thus be introduced into the
quantized Hall system if we simply view QHE as the same relative motion
between the flux quantum and the electric charge from another observation
point, i.e., the static vortex reference frame, see Fig.1(c), instead of
from the static electron reference frame for the Josephson ac effect. In
other words, for a stationary electronic background as that in the
zero-field Josephson ac effect mentioned above, the relative motion of
charge and flux is realized by the traveling vortex while in a static
magnetic background of the quantum Hall system, we assume it resulted from
the propagating charge, namely, the Hall current that travels through static
flux quanta.

From this common phase slippage scenario, one finds that the electron fluid
flowing along the $x$ direction in a planar Hall bar lying on the $x-y$
plane would build up, along the $y$ direction across the flux, a voltage
drop $V_{H}$ together with a time evolution of the phase difference $\varphi
_{12}$, which satisfy a Josephson-like voltage-frequency relation 
\begin{equation}
V_{H}=\left( \frac{\hbar }{e}\right) \omega _{JH},  \label{VH}
\end{equation}
where the ''Josephson-Hall frequency'' $\omega _{JH}=d\varphi _{12}/dt$.
With the spin-polarized electron playing the role of the superconducting
Cooper pair, this equation of the quantum Hall system can be regarded as a
single-charged version of the Josephson relation established in a
self-organized Josephson-like extended junction with the ''virtual
electrodes'' separated by a linear channel lined up along $x$ direction with
parallel quantum fluxes. According to this picture, one first see that
keeping the applied Hall current unchanged, we can increase the Hall voltage
by two ways. One is to increase the density of fluxes or the applied
magnetic field, the other is to decrease the charge carrier concentration or
the device gate voltage. Since both of the two operations increase the
number of fluxes passed, and thus the phase difference $\varphi _{12}$
experienced, by each propagating carrier within a certain period of time.
Such a continuous variation of the voltage would correspond to the ac
Josephson effect without external driving. However, in addition to the
current-induced phase evolution of the Josephson vortex, the coherent Hall
current $I_{H}=ef_{e}$ itself results in another frequency $\omega _{e}=2\pi
f_{e}$ of the traveling wave of electrons, or holes, which is also observed
at any fixed $x$ position and further provides a periodic driving for the
quantum Hall oscillation at this position. The system thus turns into a
periodically forced quantum oscillation and the quantized Hall resistance is
able to be explained by the phase-locking behavior between the
Josephson-like oscillation and the Hall current. An example with the locked
frequency ratio $2/3$, or $\nu =3/2$, is explicitly illustrated in Fig.2.

Although we regard the pictorial argument to be rather straightforward, the
corresponding theoretical formulism turns out to be apparently
unconventional. Instead of traditionally quantizing the electron motion
under external field, one follows the Josephson-RCSJ approach to construct a
semi-classical model for the Josephson-Hall oscillation which is modulated
by electron waves. The phenomenological current conservation law in $y$
direction at a fixed $x$ position should thus appear like: 
\begin{equation}
C\frac{\hbar }{e}\frac{d^{2}\varphi _{JH}}{dt^{2}}+\frac{1}{R}\frac{\hbar }{e%
}\frac{d\varphi _{JH}}{dt}+I_{c}g(\varphi _{JH})=\left| I_{e}\right|
f(\omega _{e}t),  \label{RCSJH}
\end{equation}
where $\varphi _{JH}$ represents the ''Josephson-Hall phase'' which is also
the electronic phase difference $\varphi _{12}$ across the flux channel, the
capacitance $C$ and the resistance $R$ are macroscopic parameters, $\left|
I_{e}\right| f(\omega _{e}t)$ represents the periodic driving provided by
the coherent Hall current, $I_{c}g(\varphi _{H})$ represents a nonlinear
vortex current for which a simple image can be obtained if we designate $%
g(\varphi _{H})=\sin \varphi _{H}$ to be the $y$ component of a coherent
circularly orbiting current within the quantized vortex. The Hall resistance 
\begin{equation}
R_{H}=\frac{V_{H}}{I_{H}}=\frac{\left( \frac{\hbar }{e}\right) \omega _{H}}{%
ef_{e}}=(\frac{\omega _{H}}{\omega _{e}})\frac{h}{e^{2}}  \label{RH}
\end{equation}
can thus be shown to be quantized by solving (\ref{RCSJH}) to find a locked
time-averaged frequency ratio $<\frac{\Delta \varphi _{H}}{\Delta t}>/\omega
_{e}$.

From this simple dynamical approach, it has been shown that the quantum
phase-locking scenario provides a unified mechanism for IQHE and FQHE.
Experimental supports of this semi-classical model can be found in the
reported chaotic phenomenon\cite{chaos} and a remarkable bistability with
both dissipative and non-dissipative states\cite{bistable} in the
current-induced breakdown of the QHE, these typical nonlinear dissipative
dynamical features strongly suggest a ''chaotic'' point of view for
physicists to deal with the QHE breakdown problem. Also to be noted, the
striking discovery of the novel $1/2$ state in the multilayer system\cite
{bilayer1}\cite{bilayer2} can be intuitively understood within the
phase-locking scenario as that the extra layers provide additional coupling
to tame the previously un-locked $1/2$ state.

Further study is supposed to be twofold: Experimental justification and
refinements of the equation of motion (\ref{RCSJH}), as well as its
comparison to the existing Josephson effect counterpart can do much helps to
tackle crucial open questions of QHE. For example, the sample-dependent
occurrence of the even-denominator states is expected to find its dynamical
explanation in the ''Arnol'd tongue'' pattern\cite{arnold} of the parameter
space. The current and voltage distribution, whether in the edge or the bulk
region, should be treated as a synchronized Josephson junction array issue,
no matter for the integer or the fractional quantized Hall states. The Farey
series order of appearance of the quantized plateaux and the phase
transition behaviors between the plateaux should also be compared between
these two system under a unified dynamical framework. Finally, the very
question remained to be answered turns out to be highly fundamental, if not
been taken as philosophical. As we followed the Josephson-RCSJ method to
alternatively propose an electron-perturbed semi-classical equation of
motion, we have been forced only to examine this formulism as a new
macroscopic quantization rule, or in a practical sense, a semi-classical
dual quantum theory. In either way, we are obliged to see further extension
of this temporal system to a spatial-temporal mathematical structure with a
quantum theoretical point of view. Foreseeable research should first include
a unified dynamical study of the quantum soliton in both system. Among other
quantum mechanical features, a relativistic Lorentz Contraction\cite{lorentz}
in the QHE may be discovered as the charge-flux relative velocity approaches
the sample light velocity.

In summary, adopting the hydrodynamical phase-slippage geometry together
with the semi-classical Josephson-RCSJ dynamical formulism, this letter
proposes a pictorial analogy between the periodically driven Josephson ac
effect and the quantum Hall effect. These two macroscopic quantum effects
are postulated to be both originated from the linear vortex-electron
relative motion, with the moving and standing quanta been altered from the
flux of the former to the electron of the latter. The quantized Hall
resistance of both IQHE and FQHE is suggested to be described by a unified
phase-locking dynamics of a Josephson-like oscillation periodically driven
by the electron wave of the coherent Hall current, with the time-averaged
frequency ratio playing the role of the filling factor reciprocal.

\acknowledgments
I thank Prof. Y.S. Gou, Prof. C.S. Chu, Prof. B. Rosenstein and Prof. C.C.
Chi for stimulating and helpful discussions. This work is partly supported
by the National Science Council of R.O.C. under grant No. NSC
87-2112-M009-017.

\begin{figure}[h]
\caption{ (a) An isolated vortex with equal-phase lines. (b) For a
non-propagating electron background as the ac Josephson effect, the $x$
direction relative motion between flux and charge comes from the
translational motion of the vortex. During the time $t_{f}-t_{i}$, the
vortex travels from $x_{1}$ to $x_{2}$ and the phase difference between side
1 and side 2 $\varphi _{12} $ experiences a $2\pi $ increment. (c) For a
static magnetic background as the quantum Hall effect, the relative motion
comes from the propagation of electrons. For a corresponding example, during
the time $t_{f}-t_{i}$, the electron fluid flows from $x_{2}$ to $x_{1}$ and
the phase difference between 1 and 2 is also increased by $2\pi $.}
\end{figure}

\begin{figure}[h]
\caption{ A one dimensional coherent electron wave passing through quantized
fluxes makes a periodically driven Josephson-like oscillation. Take $\beta
=3/2$ locked state as an example: During the time $\Delta t$, electron $%
e_{1} $ originally at $x_{1}$ passes through two fluxes $f_{1}$ and $f_{2}$
and causes a $4\pi $ increment for the phase difference between side 1 and
side 2, which can be measured at any certain $x$ position including $x_{2}$.
Meanwhile, there are three electron wave packets $e_{1},e_{2}$ and $e_{3}$
denoted by the sine curve, passing through $x_{2}$. Thus a phase-locked
electron-driven Josephson-like oscillation results in at $x_{2}$ the
quantization of the Hall resistance $R_{H}=V_{H}/I_{H}=(\hbar /e)(4\pi
/\Delta t)/(3e/\Delta t)=(2/3)(h/e^{2})$. Also can be seen in this figure,
the phase-locking dynamics corresponds to an elastic deformation of the flux
chain, i.e., the line density of the fluxes can be self-adjusted to maintain
the locked density ratio of electrons to fluxes.}
\end{figure}

\end{document}